\documentclass[%
 reprint,
 superscriptaddress,
 amsmath,amssymb,
 aps,longbibliography
]{revtex4-2}

\usepackage{graphicx}
\usepackage{dcolumn}
\usepackage{bm}
\usepackage{float}
\usepackage{hyperref}
\usepackage{xcolor}
\usepackage{lineno}
\usepackage{subfigure}
\usepackage[normalem]{ulem} 
\usepackage{comment}

\usepackage{xcolor}

\begin{document}

\preprint{APS/123-QED}

\title{First constraint on atmospheric millicharged particles with the LUX-ZEPLIN experiment}

\author{J.~Aalbers}
\affiliation{SLAC National Accelerator Laboratory, Menlo Park, CA 94025-7015, USA}
\affiliation{Kavli Institute for Particle Astrophysics and Cosmology, Stanford University, Stanford, CA  94305-4085, USA}

\author{D.S.~Akerib}
\affiliation{SLAC National Accelerator Laboratory, Menlo Park, CA 94025-7015, USA}
\affiliation{Kavli Institute for Particle Astrophysics and Cosmology, Stanford University, Stanford, CA  94305-4085, USA}

\author{A.K.~Al Musalhi}
\affiliation{University College London (UCL), Department of Physics and Astronomy, London WC1E 6BT, UK}

\author{F.~Alder}
\affiliation{University College London (UCL), Department of Physics and Astronomy, London WC1E 6BT, UK}

\author{C.S.~Amarasinghe}
\affiliation{University of California, Santa Barbara, Department of Physics, Santa Barbara, CA 93106-9530, USA}

\author{A.~Ames}
\affiliation{SLAC National Accelerator Laboratory, Menlo Park, CA 94025-7015, USA}
\affiliation{Kavli Institute for Particle Astrophysics and Cosmology, Stanford University, Stanford, CA  94305-4085, USA}

\author{T.J.~Anderson}
\affiliation{SLAC National Accelerator Laboratory, Menlo Park, CA 94025-7015, USA}
\affiliation{Kavli Institute for Particle Astrophysics and Cosmology, Stanford University, Stanford, CA  94305-4085, USA}

\author{N.~Angelides}
\affiliation{Imperial College London, Physics Department, Blackett Laboratory, London SW7 2AZ, UK}

\author{H.M.~Ara\'{u}jo}
\affiliation{Imperial College London, Physics Department, Blackett Laboratory, London SW7 2AZ, UK}

\author{J.E.~Armstrong}
\affiliation{University of Maryland, Department of Physics, College Park, MD 20742-4111, USA}

% ...

\collaboration{The LZ Collaboration}

\date{\today}

\begin{abstract}
We report on a search for millicharged particles (mCPs) produced in cosmic ray atmospheric interactions using data collected during the first science run of the LUX-ZEPLIN experiment.
The mCPs produced by two processes---meson decay and proton bremsstrahlung---are considered in this study. 
This search utilized a novel signature unique to liquid xenon (LXe) time projection chambers (TPCs), allowing sensitivity to mCPs with masses ranging from 10 to 1000 MeV/c$^2$ and fractional charges between 0.001 and 0.02 of the electron charge (\textit{e}).
With an exposure of 60 live days and a 5.5~tonne fiducial mass, we observed no significant excess over background. 
This represents the first experimental search for atmospheric mCPs and the first search for mCPs using an underground LXe experiment.
\end{abstract}

\maketitle

{\textit{Introduction}\textemdash}Millicharged particles (mCPs), denoted as $\chi$, are hypothetical particles carrying a small fractional electric charge to that of an electron $Q_\chi = \epsilon e$. 
The search for mCPs is closely connected to the research of string theory \cite{WEN1985651}, grand unification theories (GUTs) \cite{Pati:1973uk, Georgi:1974my}, and the principle of charge quantization \cite{Dirac:1931kp, Dobroliubov:1989mr} which is taken as an observation in the standard model (SM) without firm theoretical motivation. 
In their simplest theoretical form without considering ultraviolet completeness, mCPs can be incorporated into the SM as new particles carrying a small charge under $U(1)_Y$ gauge symmetry.  
Other possible origins include kinetic mixing between a dark photon field and the SM hypercharge field \cite{Holdom:1985ag, Holdom:1986eq, Foot:1991kb}, or extensions to the SM involving mass mixing \cite{Kors:2004dx, Cheung:2007ut, Feldman:2007wj}.
Consequently, experimental searches for mCPs not only serve as a powerful test of various dark sector models and charge quantization but also represent an important frontier in the exploration of physics beyond the SM.

Several studies \cite{PhysRevD.41.1067, Feng:2009mn, Cline:2012is} have proposed that mCPs could constitute a small portion of the dark matter (DM) in the universe, building upon the hypothesis that some fraction of DM may exhibit small electromagnetic interactions with photons \cite{Barkana:2018lgd, Pospelov:2000bq, Raby:1987ga, Sigurdson:2004zp, Antipin:2015xia}.
This interest in mCPs was notably revitalized following the anomalous findings from the EDGES experiment in 21-cm cosmology \cite{Boyarsky:2019fgp, Barkana:2018lgd, LiQiaoDan:2021ovd}, with mCPs as a potential explanation \cite{Munoz:2018pzp, Berlin:2018sjs, Kovetz:2018zan, Creque-Sarbinowski:2019mcm, Barkana:2018qrx}. 
Additionally, experimental searches for the electromagnetic interaction of galactic DM in the mass range of a few GeV/c$^2$ to~TeV/c$^2$ have been conducted to probe their interactions with nuclei \cite{PandaX:2023toi, DEAP:2020iwi, PICO:2022ohk}. 
However, for non-relativistic mCPs with smaller masses, the energy deposited during scattering with nuclei becomes small, rendering such searches challenging.

For mCPs with mass in the region 10 to 1500 MeV/c$^2$, the current experimental limits come from accelerator experiments, including  ArgoNeuT~\cite{ArgoNeuT:2019ckq}, milliQan~\cite{milliQan:2021lne}, SLACmq~\cite{Prinz:1998ua} and SENSEI~\cite{SENSEI:2023gie}.
Recent studies suggest that in this mass range, a substantial flux of relativistic mCPs may be produced through atmospheric cosmic ray interactions, generating detectable energy deposits in terrestrial detectors \cite{Harnik:2020ugb, Plestid:2020kdm, Kachelriess:2021man, ArguellesDelgado:2021lek, Du:2022hms, Du:2023hsv, Alvey:2019zaa, Wu:2024iqm}. References~\cite{ArguellesDelgado:2021lek, Harnik:2020ugb, Pokrandt:2021ptb, PandaX:2023tfq} show that liquid xenon (LXe) detectors, albeit smaller than neutrino detectors in size, can set competitive constraints on relativistic atmospheric DM particles, including mCPs \cite{Alvey:2019zaa, PandaX:2023tfq}.
The sensitivity of LXe detectors is due to their comparably low energy threshold and background rates \cite{LZ:2022ysc}.
In this letter, we report on the first search for atmospheric mCPs. 
This analysis used data collected between December 23, 2021, and May 11, 2022, from the LUX-ZEPLIN (LZ) dark matter experiment during Science Run 1 (SR1), with a total exposure of 0.91 tonne$\times$year \cite{LZ:2022lsv}.

The LZ experiment~\cite{LZ:2019sgr, Mount:2017qzi} is located 4850~ft underground in the Davis Cavern at the Sanford Underground Research Facility (SURF) in Lead, South Dakota. 
LZ has leading sensitivity to Weakly Interacting Massive Particle (WIMP) dark matter models~\cite{LZ:2022lsv, LZCollaboration:2024lux}.
At the core of the LZ experiment is a dual-phase xenon time projection chamber (TPC), a vertical cylinder approximately 1.5 meters in diameter and height, containing a 7-tonne active mass.
The TPC detects energy depositions in LXe, producing two types of signals: vacuum ultraviolet (VUV) scintillation photons (S1) and ionization electrons. The ionization electrons drift under a near-uniform electric field towards the liquid-gas surface, where they are extracted into the gas region and produce an electroluminescence signal (S2). The S1 and S2 signals are detected by two arrays of 3-inch photomultiplier tubes (PMTs), with 253 PMTs positioned at the top and 241 PMTs at the base of the TPC. The horizontal position $(x, y)$ of an event is reconstructed using the S2 light incident on the top PMT array, and the $z$-position is calculated from the delay time between S1 and S2 signals. Spatial variations in S1 and S2 signals are corrected using radioactive sources, as in~\cite{LZ:2022lsv}, yielding position-independent signals, S1\textit{c} and S2\textit{c}. The ratio of S2\textit{c} to S1\textit{c} signals is critical for differentiating between nuclear recoil (NR) and electron recoil (ER) interactions. Encasing the TPC, a `skin' of instrumented LXe and a 17-tonne gadolinium-loaded liquid scintillator outer detector (OD) serve as anti-coincidence detectors, providing shielding and background veto capabilities. Additionally, a 238-tonne ultra-pure water tank surrounds the LZ setup, further enhancing protection against ambient radioactive backgrounds.

\begin{figure}
    \centering
    \includegraphics[width=0.45\textwidth]{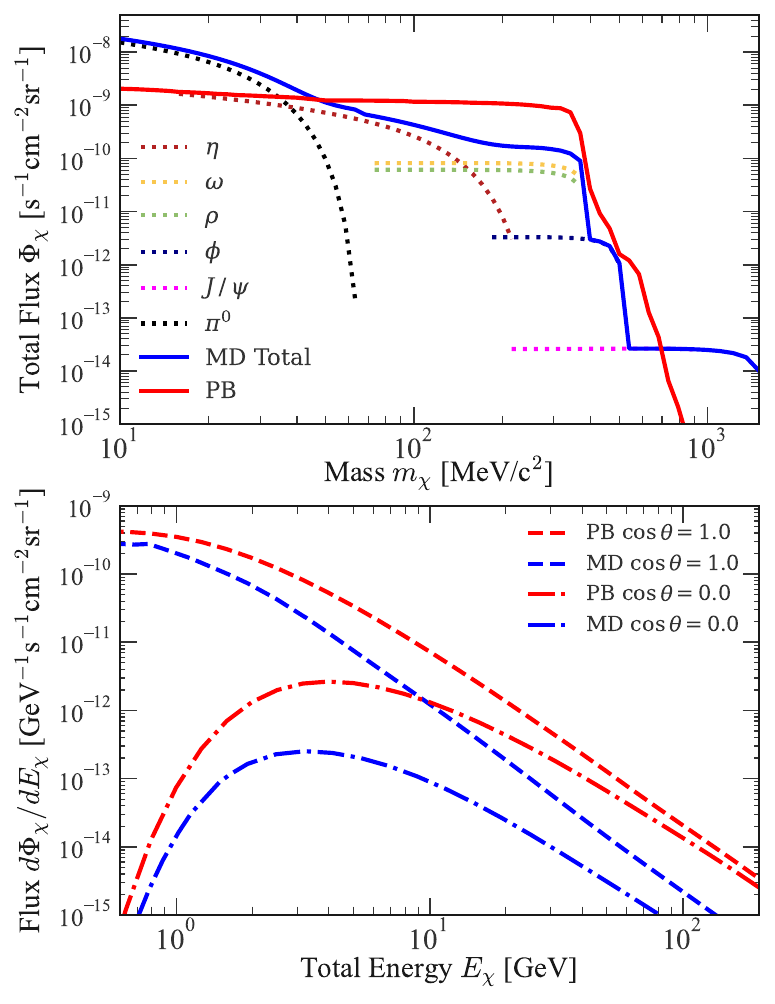}
    \caption{
    Top panel: the mCP flux integrated over all energies from the PB channel and the MD channel at Earth surface level, as a function of mCP mass $m_\chi$, with $\epsilon$ = 0.01 and  $\cos\theta = 1.0$.
    The contributions from each parent meson species to the MD channel are also shown.
    Bottom panel: the differential mCP energy spectra from the MD and PB channel reaching the LZ detector. 
    As a benchmark, we show the spectra of the mCP model with $m_\chi = 100$ MeV/c$^2$ and $\epsilon$ = 0.01.
    The underground flux from the vertical direction ($\cos\theta$ = 1.0) is represented by the solid lines, and the flux from the horizontal direction ($\cos\theta$ = 0.0) is represented by the dot-dashed lines.
    }
    \label{fig:Flux}
\end{figure}

\textit{mCP production and attenuation}\textemdash As mCPs participate in SM QED processes, the interaction between cosmic rays and atoms in the atmosphere can produce a flux of mCPs detectable by terrestrial detectors.
Following Ref.~\cite{Wu:2024iqm}, we considered mCPs produced in two distinct atmospheric production processes in this analysis: meson decay (MD) and proton bremsstrahlung (PB).
The Drell-Yan process was not considered due to its negligible flux contribution.
In the MD channel, neutral mesons are produced during cosmic ray hadronic interactions, leading to the generation of mCP-pairs through electromagnetic decays. 
The mCP flux from MD can be calculated with a zenith angle dependence using the cascade equation as in Ref.~\cite{Gondolo:1995fq}. 
For this calculation, we employed the \textsc{HeavenlyMCP} package \cite{ArguellesDelgado:2021lek} which estimates the mCP flux at the Earth's surface from MD, incorporating the meson-mCP branching ratio as detailed in Refs.~\cite{Harnik:2020ugb, Plestid:2020kdm, Gorbunov:2021jog}. 
The uncertainty on the mCP flux in the MD channel, arising from cosmic ray and hadronic interaction models, was evaluated to range from 26\% to 71\% depending on the mediating meson type~\cite{ArguellesDelgado:2021lek}.
In the PB channel, a cosmic ray proton collides with atmospheric atoms and is stopped. 
The initial state proton emits a bremsstrahlung photon, which subsequently radiates a mCP pair \cite{Gninenko:2018ter, Fermi:1924tc, Williams:1934ad, vonWeizsacker:1934nji}.
The atmospheric mCP flux at the surface from PB was adopted from the calculations presented in Refs.~\cite{Du:2022hms, Wu:2024iqm}.
The uncertainty on the mCP flux produced in PB mainly arises from the proton off-shell form factor, 
which leads to an average 46$\%$ flux uncertainty in the mass range of interest, 
derived from the calculations in Ref.~\cite{Du:2022hms}.
The mCP flux at the surface of the Earth produced from both processes is illustrated in the top panel of Fig.~\ref{fig:Flux}.

Atmospheric mCPs lose energy through scattering and ionization as they traverse the overburden on their way to an underground detector.
This attenuation effect will reduce the flux reaching the LZ detector, and scales with the charge fraction $\epsilon$ and distance traveled through the Earth.
To account for this, we employed the energy-loss based attenuation method outlined in Ref.~\cite{ArguellesDelgado:2021lek} to calculate the attenuated mCP flux reaching the LZ detector, starting from the initial surface flux, with the SURF surface profile as in Ref.~\cite{LZ:2020zog} taken into consideration.
For illustration, the mCP flux produced in both channels reaching the LZ detector from zenith angle $\cos\theta=1.0$ and $0.0$ is shown in the bottom panel of Fig.\ \ref{fig:Flux}.

\textit{mCP Signal Modeling in LZ}\textemdash We simulated mCP tracks through the LZ geometry using the underground mCP flux shown in Fig.~\ref{fig:Flux} for different incoming angles, assuming no deflection and instantaneous traversal due to their high kinetic energy. The mCPs produced in cosmic ray interactions are highly relativistic, typically crossing the LZ TPC in a few nanoseconds. Energy depositions along these tracks were sampled using the photon absorption ionization (PAI) model~\cite{Allison:1980vw} (see Appendix~A for details and an alternative approach, the free electron model).
The LZ simulation framework~\cite{LZ:2020zog}, incorporating \textsc{\footnotesize NEST} 2.3.12~\cite{nest:2.3.12}, was employed to characterize the detector response, converting energy depositions into S1$c$, S2$c$, drift time, S2 width, and other observables. 
Since mCPs interact with xenon similarly to $\beta$ particles, we adopted the \textsc{\footnotesize NEST} $\beta$-like ER yield model.
The low-energy ER yield model in \textsc{\footnotesize NEST} is constrained by experimental data from Refs.~\cite{LUX:2017ojt, Boulton:2017hub} and further fine-tuned using LZ \textit{in situ} tritium $\beta$ calibration data.

\begin{figure}[!htbp]
    \centering
    \includegraphics[width=1\linewidth]{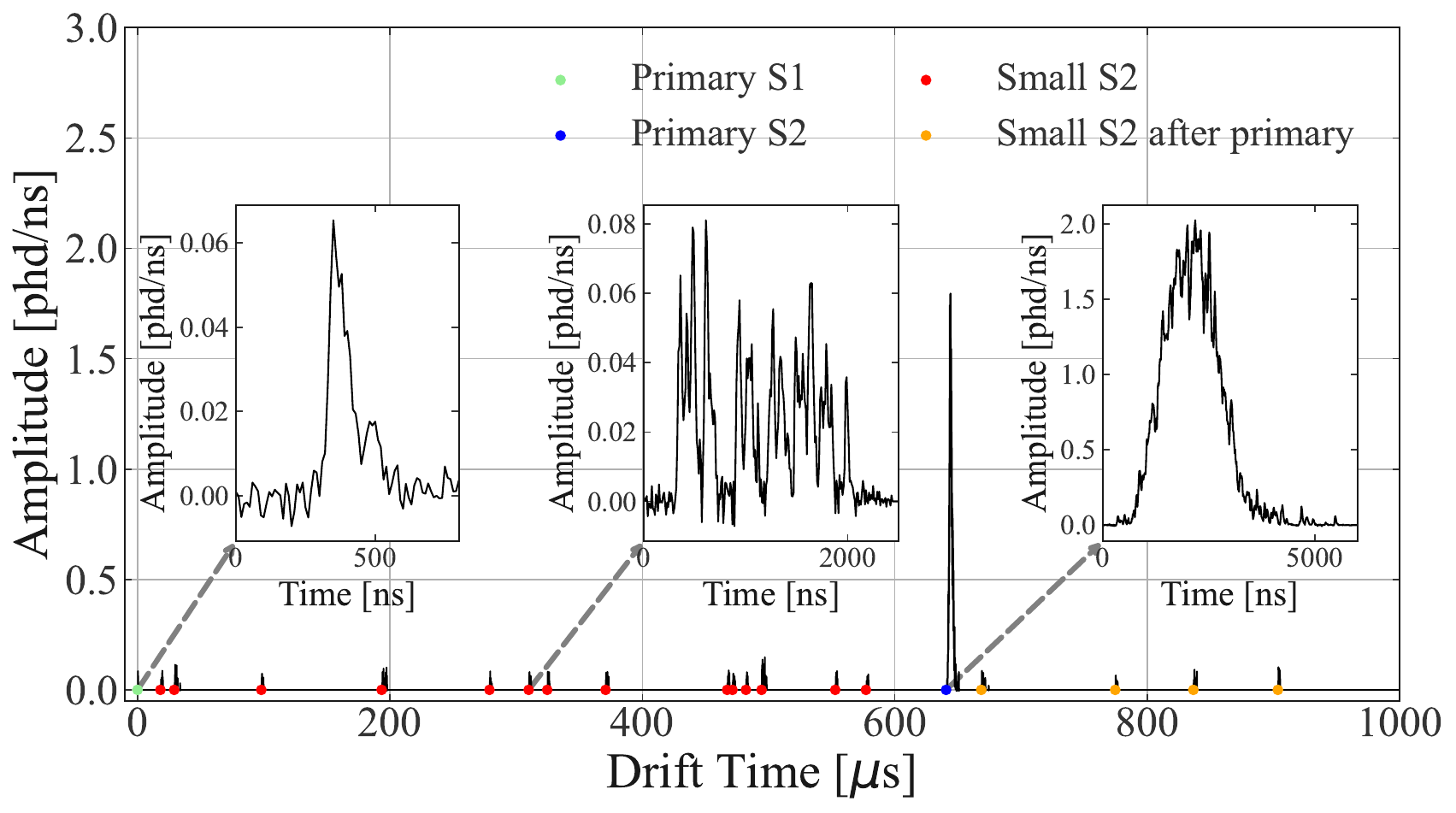}
    \caption{
    An example event topology of a simulated mCP event in the LZ TPC, with mCP mass $m_\chi$ = 0.1 GeV/$c^2$, and $\epsilon=0.003$.
    }
    \label{fig:topology}

\end{figure}

Simulations indicate that mCPs with $\epsilon > 0.001$ undergo multiple scatterings along their tracks, though only a small fraction of these scatterings result in hard scatters with energy depositions $\geq 1$~keV, sufficient to produce a detectable S1-S2 pair~\cite{LZ:2023poo}.
For instance, in the PAI model, an mCP with $m_\chi = 100$ MeV/c$^2$, $\beta\gamma = 4$, and $\epsilon = 0.003$ has a mean free path of 54 mm, much smaller than the dimensions of the LZ TPC. 
However, most scatterings are soft~($< 1$ keV) collisions, producing only small S2 pulses, including single electrons~(SE). 
The mean free path for a hard scatter of the same mCP is approximately 2.6~m, comparable to the size of the LZ TPC. Hence, about 32\% of mCPs crossing the TPC produce a single hard scatter (SHS), while 11\% produce multiple hard scatters (MHS).
As $\epsilon$ increases, the fraction of MHS also increases. 
Consequently, a typical mCP traversing the TPC produces a sequence of soft scatters and one or more hard scatters. 
All S1 pulses from different scatters merge, while S2 pulses—having microsecond-scale widths—become unresolvable if scatters occur at similar depths. 
The simulation accounts for pulse merging by modeling time separation, pulse widths, and pulse areas~\cite{LZ:2020zog, Allison:2016lfl}. 
The simulated topology and waveform of a typical mCP SHS event for $\epsilon = 0.003$ is shown in Fig.~\ref{fig:topology}. This distinct event topology is characteristic of mCP interactions and enables LZ to probe mCPs with charge fractions between 0.001 and 0.02.

\textit{Event selection and efficiency}\textemdash
To identify mCP signals, we restricted our search to events with SHS producing a primary S1-S2 pair, leveraging the well-established SR1 WIMP search analysis, which utilizes single-scatter (SS) events~\cite{LZ:2022lsv}. Additionally, selected events must contain at least three small S2s between the primary S1 and S2, produced by soft scatters.
The small S2s after the primary S2 are not considered due to elevated rates of activity after large S2s, as is commonly observed in dual-phase xenon TPCs~\cite{LUX:2020vbj, Sorensen:2017ymt}.
The S1 signal requires a threefold PMT coincidence and an S1\textit{c} of at least 3~phd.  
The S2 threshold is set at S2\textit{c} = 2000~phd, corresponding to $\sim 1$~keV. S2 signals above this threshold are considered primary, while those below are classified as small S2s.
The primary S1 and S2 signals are required to pass the same data quality event selection criteria as in the SR1 WIMP search, except for a prompt OD veto cut and an `excess area' cut~\cite{LZ:2022lsv}. 
The prompt OD veto cut removes events with coincident signals in the OD within 300~ns of the TPC signal, and the `excess area' cut rejects events where the summed area of pulses between the primary S1 and S2 exceeds that of the primary S2 pulse. 
Both cuts could potentially exclude mCP signals. 
Removing these two cuts maximizes the acceptance of mCP signal events while still maintaining rejection of accidental coincidences of isolated S1 and S2 pulses from the other data quality selections.
In addition, an mCP region of interest (ROI) is defined for the primary S1 and S2 in the \{S1\textit{c}, log$_{10}$(S2\textit{c})\} observable space. This contour is derived from simulations of mCP SS events at the 90\% confidence level (CL), with energy deposits sampled using the PAI model. This ROI cut removes events in which the primary S2 arises from the misclassification of unresolved multiple energy deposits in the TPC, as well as events with nuclear recoils from background neutrons.

\begin{figure}[ht]
    \centering
    \includegraphics[width=1\linewidth]{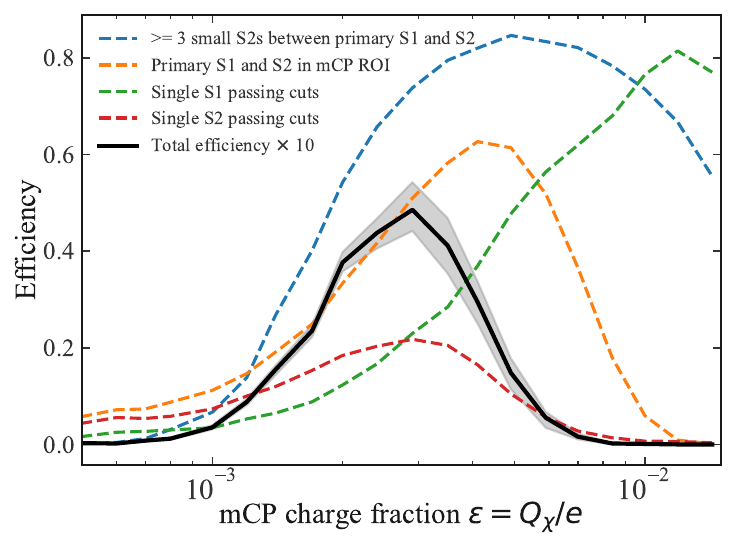}
    \caption{
    Efficiency of the data selection cuts evaluated from simulations, defined as number of tracks passing the cut divided by number of tracks simulated.
    The total efficiency has been scaled up by a factor of 10 for better visibility. The gray error band represents the total uncertainty in the efficiency.
    }
    \label{fig:Acceptance}
\end{figure}
The total signal efficiency after event selection, shown in Fig.~\ref{fig:Acceptance}, was evaluated solely from simulation, following the mCP signal modeling in LZ, and peaks at $\epsilon \simeq 0.003$. 
The error band (gray) is the systematic uncertainties quantified from different S2 width cuts, combined with the uncertainties on the hardware trigger efficiency and other data quality cuts~\cite{LZ:2022lsv}.
At lower charges, efficiency is primarily suppressed by the absence of small S2 pulses between the primary S1 and S2, as well as the lack of a hard scatter to generate the primary S1-S2 pair.
For higher charges, most of the signal loss arises from MHS, and also the primary S2 merging with other S2s, causing the event to move out of the mCP ROI or fail data quality selection.

\begin{figure}[t!]
    \centering
    \includegraphics[width=1\columnwidth]{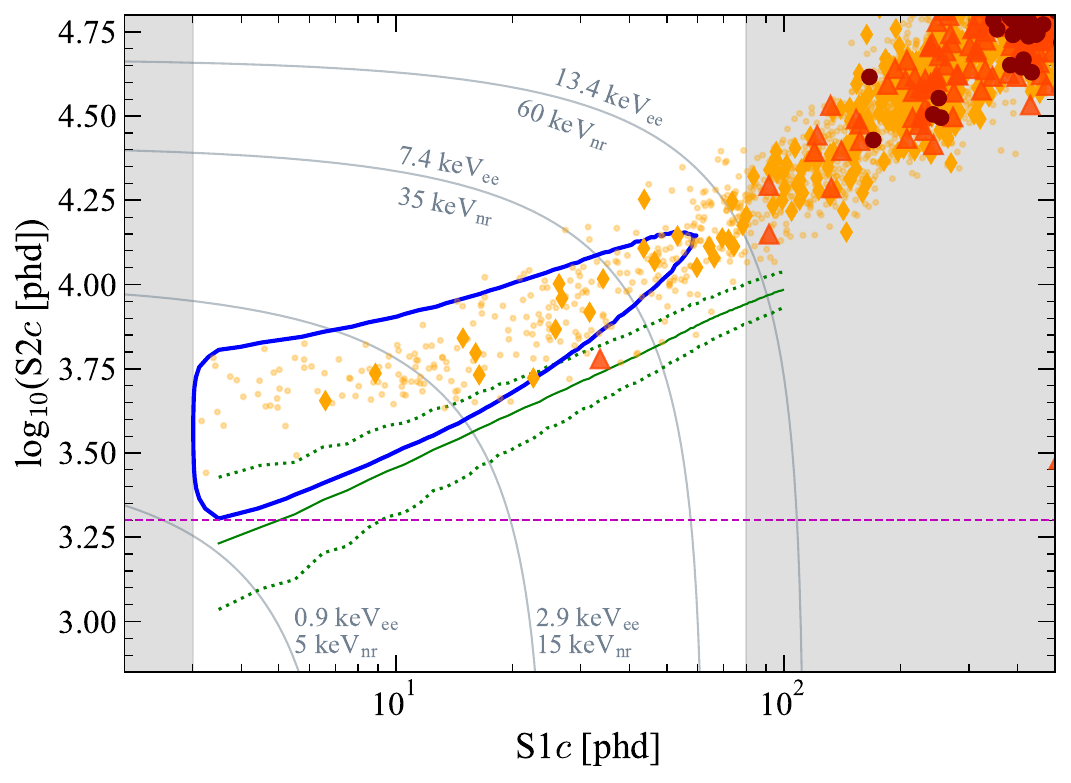}
    \caption{LZ SR1 mCP search SHS events 
    after data selections, excluding the ROI cut and before applying the three small S2 selection, shown in the log$_{10}$(S2\textit{c})-S1\textit{c} space within an extended S1 window (up to 500 phd). Events are marked by small orange points (zero small S2 pulses between the primary S1 and S2), orange diamonds (one small S2 pulse), dark-orange triangles (two small S2 pulses), and large filled dark-red circles (three or more small S2 pulses). The trend of increasing small S2 pulses at higher S1 values is due to the increased probability of S1-induced SEs from the photoionization of bulk impurities~\cite{LUX:2020vbj}. The unshaded region denotes the WIMP ROI (S1\textit{c} 3-80 phd), while solid blue contours show the mCP ROI. 
    The dashed magenta line marks the 2000 phd threshold. The solid green line represents the median of a uniform NR background, with dashed lines indicating the 10\% and 90\% quantiles. Thin gray lines outline contours of constant energy.}
    \label{fig:signal_contour}
\end{figure}

\textit{Backgrounds}\textemdash Two primary background sources could mimic mCP signal signatures. The dominant background arises from SS background events that produce a primary S1-S2 pair, accidentally coinciding with random small S2 pulses. 
The SS background was estimated from the SR1 WIMP search analysis~\cite{LZ:2022lsv, LZ:2022ysc}, whose ROI fully encompasses the mCP ROI (Fig.~\ref{fig:signal_contour}). Before requiring three or more small S2 pulses, we expect $209 \pm 22$ SS background events in the mCP ROI, accounting for the removal of the prompt OD veto and `excess area' cuts. 

To quantify the likelihood of random small S2 pulses appearing between the primary S1 and S2, we used two datasets: sideband data from a pre-S1 window (a time region before the S1) and tritium calibration data. 
The small S2 rate correlates with S1 pulse area, primarily due to S1-induced SE emissions from the photoionization of bulk impurities~\cite{LUX:2020vbj}. 
The pre-S1 window, which lacks a preceding S1, provides a lower bound on the small S2 rate, whereas tritium data, exhibiting enhanced activity between S1 and S2, offers a conservative upper bound. 
From these measurements, we estimated that 0.07\% to 0.2\% of SS background events contain three or more small S2 pulses. 
To validate this estimate, we analyzed a separate sideband dataset of events near the TPC wall, selected under conditions that mimic SS backgrounds but exclude mCP signals. 
Out of 698 events, one exhibits three or more small S2 pulses, consistent with the predicted range of [0.5, 1.4] events. 
As a further cross-check, we relaxed the criterion to require only two small S2 pulses, finding six observed events, consistent with the predicted range of [3.5, 6.9].  

The second background category arises from multiple-scatter (MS) events, where gamma rays and neutrons from detector radioactivity undergo multiple scatterings in the TPC. 
However, simulations indicate that such MS events rarely produce the mCP-like event topology, with an expected contribution of $<0.01$ events, making this background negligible.  

Combining these factors, the total expected background in the mCP ROI is [0.15, 0.42] events, obtained by multiplying the SS background expectation (209 events) by the probability of an accidental coincidence of three or more small S2 pulses ([0.07\%, 0.2\%]).

\textit{Results and discussion}\textemdash A search for mCPs was performed using LZ SR1 data, applying the same live time exclusions as those used in the WIMP search~\cite{LZ:2022lsv}. 
This resulted in an effective live time of $60\pm1$ days and a fiducial LXe mass of $5.5\pm0.2$ tonnes.
The events that pass all selections except the mCP ROI and the three small S2 cuts are depicted in Fig.~\ref{fig:signal_contour}, with the mCP ROI overlaid as a blue contour.
After all cuts were applied, no events were observed, which aligns with the expectations from our background model. 
The probability of observing zero events, given the background prediction, lies between $66\%$ and $86\%$. 

\begin{figure}
    \centering
    \includegraphics[width=1\columnwidth]{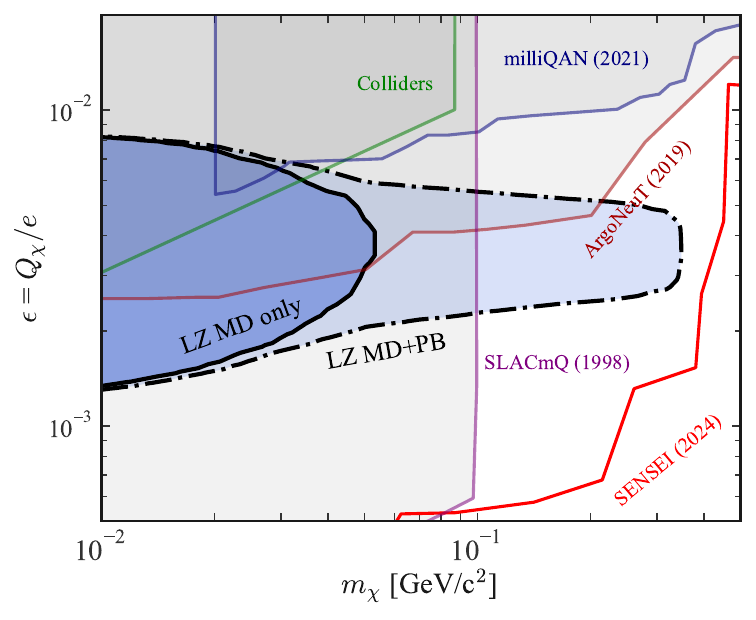}
    \caption{
    The 90$\%$ CL limits on mCP fractional charge $\epsilon = Q_\chi/e$  derived from atmospheric production channels, as a function of mCP mass $m_\chi$.
    The black solid lines are the constraints from the MD process only, and dot-dashed lines are from both the MD and PB processes combined.
    Selected limits on mCPs from beam experiments are also shown \cite{ArgoNeuT:2019ckq, milliQan:2021lne, Davidson:2000hf, Prinz:1998ua, SENSEI:2023gie}. }
    \label{fig:PLRResult}
\end{figure}

We derived atmospheric mCP exclusion limits following the Feldman-Cousins \cite{Feldman:1997qc} procedure for a zero observed signal scenario. 
To obtain a conservative constraint, we set the background expectation to 0.15 events. 
Atmospheric mCP models yielding more than 2.26 signal events, after accounting for the detection efficiency as a weight factor in the calculation, are excluded at the 90$\%$ confidence level.
The results are shown in Fig.\ \ref{fig:PLRResult} along with other experimental limits from  beam experiments, including ArgoNeuT \cite{ArgoNeuT:2019ckq}, milliQan demonstrator \cite{milliQan:2021lne}, colliders \cite{Davidson:2000hf}, SLACmq \cite{Prinz:1998ua} and SENSEI \cite{SENSEI:2023gie}. 
After combining contributions from the MD and PB processes, the lower bound of the LZ constraint on atmospheric mCPs 
is at $\epsilon\sim 0.002$ for mCP masses < 300 MeV/c$^2$, where the surface mCP flux begins to drop sharply.
The closed-contour shape of the constraints is from the fact that the selection efficiency drops to zero for $\epsilon < 0.001$ and $\epsilon > 0.01$, as shown in Fig.~\ref{fig:Acceptance}.
We estimated a 11$\%$ uncertainty on the constraints from flux and data selection efficiency uncertainties.
\vspace{0pt}

\textit{Conclusions}\textemdash We have presented the inaugural experimental search for mCPs produced in cosmic ray atmospheric interactions, which is highly complementary to existing results from accelerator-based experiments.
We have considered mCPs originating from two production channels: meson decay and proton bremsstrahlung, and conducted the search using a signature novel to liquid xenon detectors.
Utilizing data from LZ SR1, we found the data to be consistent with the background-only hypothesis for all tested mCP model parameters, and based on this result, we set the first constraints on atmospheric mCP models.

\textit{Acknowledgments}\textemdash We thank Zuowei Liu, Rundong Fang, Mingxuan Du and Zhen Liu for the PB process flux and useful discussions. The research supporting this work took place in part at the Sanford Underground Research Facility (SURF) in Lead, South Dakota. Funding for this work is supported by the U.S. Department of Energy, Office of Science, Office of High Energy Physics under Contract Numbers DE-AC02-05CH11231, DE-SC0020216, DE-SC0012704, DE-SC0010010, DE-AC02-07CH11359, DE-SC0015910, DE-SC0014223, DE-SC0010813, DE-SC0009999, DE-NA0003180, DE-SC0011702, DE-SC0010072, DE-SC0006605, DE-SC0008475, DE-SC0019193, DE-FG02-10ER46709, UW PRJ82AJ, DE-SC0013542, DE-AC02-76SF00515, DE-SC0018982, DE-SC0019066, DE-SC0015535, DE-SC0019319, DE-SC0024225, DE-SC0024114, DE-AC52-07NA27344, \& DE-SC0012447. This research was also supported by U.S. National Science Foundation (NSF); the UKRI’s Science \& Technology Facilities Council under award numbers ST/W000490/1, ST/W000482/1, ST/W000636/1, ST/W000466/1, ST/W000628/1, ST/W000555/1, ST/W000547/1, ST/W00058X/1, ST/X508263/1, ST/V506862/1, ST/X508561/1, ST/V507040/1, ST/W507787/1, ST/R003181/1, ST/R003181/2,  ST/W507957/1, ST/X005984/1, ST/X006050/1; Portuguese Foundation for Science and Technology (FCT) under award numbers PTDC/FIS-PAR/2831/2020; the Institute for Basic Science, Korea (budget number IBS-R016-D1); the Swiss National Science Foundation (SNSF)  under award number 10001549. This research was supported by the Australian Government through the Australian Research Council Centre of Excellence for Dark Matter Particle Physics under award number CE200100008. We acknowledge additional support from the UK Science \& Technology Facilities Council (STFC) for PhD studentships and the STFC Boulby Underground Laboratory in the U.K., the GridPP~\cite{faulkner2005gridpp,britton2009gridpp} and IRIS Collaborations, in particular at Imperial College London and additional support by the University College London (UCL) Cosmoparticle Initiative, and the University of Zurich. We acknowledge additional support from the Center for the Fundamental Physics of the Universe, Brown University. K.T. Lesko acknowledges the support of Brasenose College and Oxford University. The LZ Collaboration acknowledges the key contributions of Dr. Sidney Cahn, Yale University, in the production of calibration sources. This research used resources of the National Energy Research Scientific Computing Center, a DOE Office of Science User Facility supported by the Office of Science of the U.S. Department of Energy under Contract No. DE-AC02-05CH11231. We gratefully acknowledge support from GitLab through its GitLab for Education Program. The University of Edinburgh is a charitable body, registered in Scotland, with the registration number SC005336. The assistance of SURF and its personnel in providing physical access and general logistical and technical support is acknowledged. We acknowledge the South Dakota Governor's office, the South Dakota Community Foundation, the South Dakota State University Foundation, and the University of South Dakota Foundation for use of xenon. We also acknowledge the University of Alabama for providing xenon. For the purpose of open access, the authors have applied a Creative Commons Attribution (CC BY) license to any Author Accepted Manuscript version arising from this submission. Finally, we respectfully acknowledge that we are on the traditional land of Indigenous American peoples and honor their rich cultural heritage and enduring contributions. Their deep connection to this land and their resilience and wisdom continue to inspire and enrich our community. We commit to learning from and supporting their effort as original stewards of this land and to preserve their cultures and rights for a more inclusive and sustainable future.


\bibliography{main}

\section*{End Matter}
\appendix

\begin{figure}[t]
    \centering
    \includegraphics[width=0.5\textwidth]{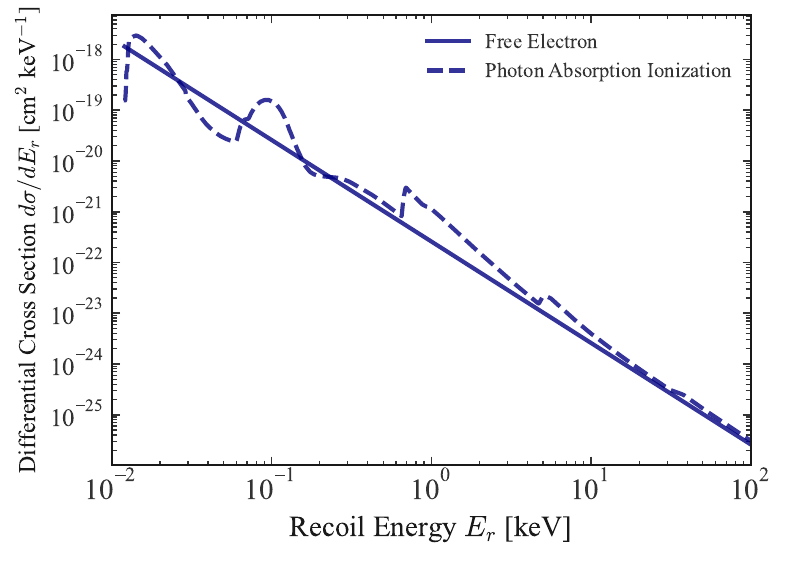}
    \caption{
    The differential cross section of mCP-electron interaction in liquid xenon predicted by the free electron model (solid) and the PAI model (dashed). 
    As a benchmark, we take $m_\chi$ = 100 MeV/c$^2$, $E_\chi = 0.5$ GeV and $\epsilon=1$ .
    }
    \label{fig:XSection}
\end{figure}

\begin{figure}[t]
    \centering
    \includegraphics[width=1\linewidth]{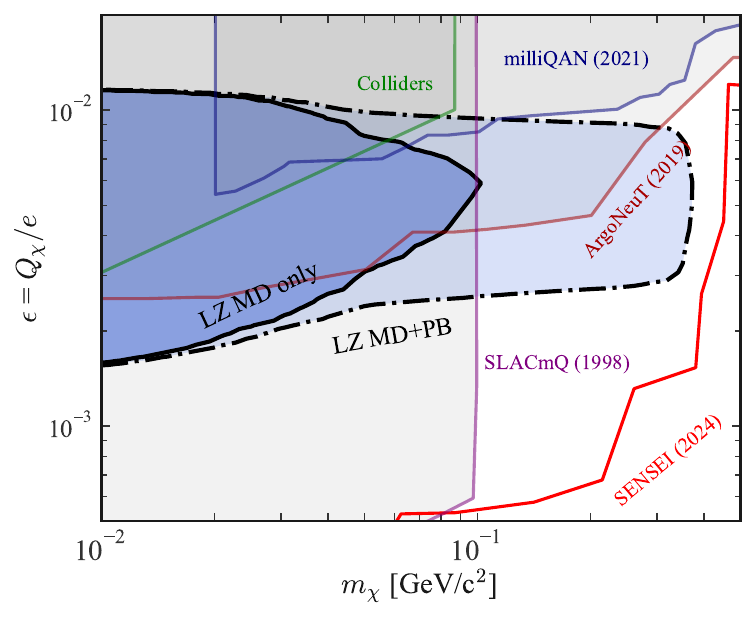}
    \caption{The 90$\%$ CL limits on mCP fractional charge $\epsilon = Q_\chi/e$ derived from atmospheric production channels, as a function of mCP mass $m_\chi$, using the free electron interaction model.
    The black solid lines are the constraints from the MD process only, and dot-dashed lines are from both the MD and PB processes combined.
    Selected limits on mCPs from beam experiments are also shown \cite{ArgoNeuT:2019ckq, milliQan:2021lne, Davidson:2000hf, Prinz:1998ua, SENSEI:2023gie}.}
    \label{fig:PLRResultFree}
\end{figure}

{\textit{Appendix A: mCP interaction models}\textemdash}In this section, we discuss different mCP interaction models in liquid xenon.  
The PAI model~\cite{Allison:1980vw}, used in our main results, accounts for xenon's electron binding energies using optical constants and has been employed in previous mCP searches~\cite{SuperCDMS:2020hcc,CONNIE:2024off}. Following Ref.~\cite{BihariPrasad:2013mfw}, the optical constants used in our calculations are taken from Refs.~\cite{CXROdatabase, Palik:396087}.
The mCP-electron scattering differential cross section predicted by the PAI model is related to that of muons as
\begin{equation}
\frac{d\sigma^\text{PAI}_\chi(\beta)}{dE_r} = \epsilon^2\frac{d\sigma^\text{PAI}_\mu(\beta)}{dE_r},
\end{equation}
where $\beta = v/c$ is the velocity of the ionizing particle, and $\sigma_\mu^\text{PAI}$ is the PAI cross section for muons.

The free electron model~\cite{Harnik:2020ugb, Plestid:2020kdm}, which assumes all electrons in LXe are free, is widely used in recent mCP theory calculations~\cite{ArguellesDelgado:2021lek, Du:2022hms, Harnik:2020ugb}. The mCP-free electron scattering differential cross section is given by \cite{Plestid:2020kdm, Plestid:2020kdm}:
\begin{equation}
    \frac{d\sigma^\text{FE}}{dE_r}=\epsilon^2 \alpha^2 \frac{E_r + 2E^2_\chi/E_r - 2E_\chi - m_e - m_\chi^2/m_e}{E_r m_e (E^2_\chi - m^2_\chi)},
\end{equation}
where $\alpha$ is the fine structure constant, $E_r$ is the recoil energy, $m_\chi$ is the mass of the mCP and $m_e$ is the mass of an electron.

The differential cross section derived from both models are shown in Fig.~\ref{fig:XSection}.
We note that the cross section is approximately inversely proportional to the square of the electron recoil energy, $E_r$. 
Due to the xenon electron shell structure, the PAI model predicts peaks near 10, 100, and 1000 eV, leading to an increased number of scattering events at these energies, which are not seen in the free electron model.

We conducted the analysis separately using the free electron model.
The excluded region derived from the free electron model is shown in Fig.~\ref{fig:PLRResultFree}.
Compared to the excluded region derived from the PAI model, the contour sees an upward shift, due to the smaller cross section, and the lack of peaks near 1 keV recoil energy.

\textit{Appendix B: Data availability}\textemdash Selected data from this analysis are publicly available at \href{https://tinyurl.com/LZDataReleaseSR1mCP}{https://tinyurl.com/LZDataReleaseSR1mCP}, including the following:

\textbf{Figure 4}: SR1 mCP search data after selection, provided in the format (S1\textit{c}, log$_{10}$(S2\textit{c}), small S2 count between the primary S1 and S2).

\textbf{Figure 5}: Points representing the 90$\%$ CL exclusion contours of MD-only and MD+PB combined.

\end{document}